\documentclass{sig-alternate-05-2015}

\usepackage{fixltx2e}
\usepackage{algorithm}
\usepackage[noend]{algpseudocode}
\MakeRobust{\Call}
\usepackage{tikz}
\usepackage{url}

\usepackage{etoolbox}
\makeatletter
\patchcmd{\maketitle}{\@copyrightspace}{}{}{}
\makeatother

\newcommand{\pasta}{\texttt{PaStA}}
\newcommand{\prt}{Preempt-RT}

\begin{document}
\conferenceinfo{}{}
\title{Observing Custom Software Modifications: A Quantitative Approach of Tracking the Evolution of Patch Stacks}
\numberofauthors{3}
\author {
	\alignauthor
	{
		Ralf Ramsauer\\
		\affaddr{Technical University of Applied Sciences Regensburg}\\
		\email{ralf.ramsauer@othr.de}
	}
	\alignauthor
	{
		Daniel Lohmann\\
		\affaddr{Friedrich-Alexander University Erlangen-Nuremberg}\\
		\email{lohmann@cs.fau.de}
	}
	\alignauthor 
	{
		Wolfgang Mauerer\\
	   \affaddr{Technical University of Applied Sciences Regensburg}\\
	   \affaddr{Siemens AG, Munich}\\
	   \email{wolfgang.mauerer@othr.de}
	}
}

\maketitle

\begin{abstract}
	Modifications to open-source software (OSS) are often provided in the form
	of ``patch stacks'' -- sets of changes (patches) that modify a given
	body of source code.  Maintaining patch stacks over extended periods of
	time is problematic when the underlying base project changes frequently.
	This necessitates a continuous and engineering-\hspace*{0mm}intensive
	adaptation of the stack. Nonetheless, long-term maintenance is an important
	problem for changes that are not integrated into projects, for instance
	when they are controversial or only of value to a limited group of users.
	
	We present and implement a methodology to systematically examine the
	temporal evolution of patch stacks, track non-functional properties like
	integrability and maintainability, and estimate the eventual economic
	and engineering effort required to successfully develop and maintain patch
	stacks.
	Our results provide a basis for quantitative research on patch stacks,
	including statistical analyses and other methods that lead to actionable
	advice on the construction and long-term maintenance of custom extensions to OSS.
\end{abstract}

\section{Introduction}
Special-purpose software, like industrial control, medical analysis, or other
domain-specific applications, is often composed of contributions from
general-purpose projects that provide basic building blocks.
Custom modifications implemented on top of them fulfill certain additional
requirements, while the development of \emph{mainline}, the primary branch of
the base project, proceeds independently.

Especially for software with high dependability requirements, it is crucial to
keep up to date with mainline: latest fixes must be applied and new general
features have to be introduced, as diverging software branches are hard to
maintain and lead to inflexible systems~\cite{mergeconflict}.
Parallel development often evolves in the form of \emph{patch stacks}:
feature-granular modifications of mainline releases.
Because of the dynamics exhibited by modern software projects, maintaining patch
stacks can become a significant issue in terms of effort and costs.

Our toolkit \pasta{}\footnote{\url{https://github.com/lfd/PaStA}}
(\textbf{Pa}tch \textbf{St}ack \textbf{A}nalysis) quantitatively analyses the
evolution of patch stacks by mining git~\cite{git} repositories and produces
data that can serve as input for statistical analysis.
It compares different releases of stacks and groups similar patches (patches
that lead to similar modifications) into equivalence classes.
This allows us to compare those classes against the base project to measure
integrability and influence of the patch stack on the base project.
Patches that remain on the external stack across releases are classified as
\emph{invariant} and are hypothesised to reflect the maintenance cost of the
whole stack.
A fine grained classification of different patch types that depends on the
actual modifications could function as a measure for the \emph{invasiveness} of
the stack.

In summary, we claim the following contributions:
\begin{itemize}
	\item We provide an approach and tool for observing the evolution
		  of patch stacks.  
	\item We propose a language-independent
		  semi-automatic algorithm based on string distances that is suitable
		  for detecting similar patches on patch stacks.
	\item We provide a case study on \prt{}~\cite{prt}, a realtime extension of
	      the Linux kernel that enjoys wide\-spread use in industrial appliances
		  for more than a decade, yet has not been integrated into standard Linux.
		  We measure its influence on mainline and visualise the development
		  dynamics of the stack.
\end{itemize}

\section{Approach}

In general, a patch stack (also known as patch set) is defined as a set of
patches (commits) that are developed and maintained independently of the base
project.
Well-known examples include the \prt{} Linux realtime extension, the Linux LTSI
(Long Term Support Initiative) kernel, and vendor-specific Android stacks needed
to port the system to a particular hardware.
In many cases, patch stacks are applied on top of individual releases
of an upstream version, but they do not necessarily have to be
developed in a linear way~\cite{bird09}.
The commits of the patched version of a base project are identified as the set of
commit hashes that do not occur in the mainline project.

Our analysis is based on the following assumptions:
\begin{itemize}
	\item Mainline \emph{upstream} development takes place in one single branch.
	\item Every release of the patch stack is represented by a separate branch.
\end{itemize}

The work flow of \pasta{} consists of the following steps:
(1) Set up a repository containing all releases of the patch stacks.
(2) Identify and group similar patches across different versions of the patch
   stacks.
(3) Compare representatives of those groups against mainline.
(4) Use statistical methods to draw conclusions on the development and evolution
   of the patch stacks.

A \emph{commit hash} provides a unique identifier for every commit: In
the following, $U$ is the set of all commit hashes of the base project,
while $P_i$ is the set of the commit hashes of a release $i$ of the patch stacks.
$P\equiv \bigcup_i P_i$ denotes all commit hashes on the patch stacks.
Note that $P\cap U = \emptyset$.  Let $H\equiv P \cup U$ be the set of
all commit hashes of interest.  A semi-automatic classification
function $\text{comp}: P \times H \rightarrow \{\text{True},
\text{False}\}$ decides whether two patches are similar or not.  A
detailed description of the function $\text{comp}$ can be found in
Section~\ref{sec:detect}.

In the implementation, \pasta{} mines git repositories.
Without loss on generality, we focus on
this particular version control system because it is widely employed
in current OSS development.

\subsection{Grouping Similar Patches}
Patch stacks change as they are being aligned with the changes in base
project and additionally integrate or loose functionalities.  New
patches are pushed on top of the stack, existing patches may be
amended to follow up with API changes, or patches are dropped.
Because of the rapid dynamics and growth of Open Source
projects~\cite{osgrowth}, a significant amount of patches must manually be
ported from one release of the base project to the next.  Since the
base project changes over time, it is necessary to
continuously adapt the details of individual patches.  Those
adaptations can be classified in textual and higher-order
conflicts~\cite{brun}.  Textual conflicts can be solved by manually
porting the patch to the next version.  In a series of patches,
patches may depend on each other, so that textual conflicts in one
patch lead to follow-up conflicts in further patches.  Higher-order
conflicts occur when a patch obtains a new (erroneous) semantic
meaning after changes in the base project diverged, despite a lack of
textual conflicts.  Both types are known to induce high maintenance
cost~\cite{ltsi}.

Even if the semantics of patches remain invariant over time (e.g., a
patch introduces identical functional modifications in subsequent
revisions of the patch), their textual content can change considerably
over time.
To track patches with unchanged semantics over time, we introduce the classifier
function $\text{comp}$ that places similar patches into equivalence classes
$R_j$, so that $P=\bigcup_j R_j$. If $\text{comp}$ were able to track the exact
semantics of patches, it would hold that
$\text{comp}(a,b) = \text{yes} \Leftrightarrow a \sim b$.
But as $\text{comp}$ can only compare textual changes, it follows that
$\text{comp}(a,b) = \text{yes} \Rightarrow a \sim b$.
This results from the fact that two similar patches between two successive
versions usually have less textual changes than the first and last occurrence of
the same patch. We approximate $P \approx \bigcup_j \hat{R_j}$.

\subsection{Comparing Groups Against Mainline}
After grouping all patches on the stacks in equivalence classes
$\hat{R_j}$, a complete representative system $\mathcal{R} \subseteq
P$ is chosen and compared against the commits in the base project.
As representative of an equivalence class, we choose the patch with the latest
version.
$Q = \{(r, u) | r\in \mathcal{R}, u \in U, \text{comp}(r,u)=1\}$ denotes the set
of all patches that are found in the base project.

\newpage
\subsection{Detecting Similar Patches}
\label{sec:detect}
To group patches into equivalence classes and find them in the base project, it
is necessary to detect similar commits.
Generally, a commit consists of a unique hash, a descriptive message that
informally summarises the modifications, and so called \emph{diffs}~\cite{diff}
that describe the actual changes of the code.

Existing work on detecting similar code fragments primarily targets on detecting
code duplicates~\cite{codedup} or on revealing code plagiarism.
Possible approaches include language-dependent lexical analysis, code
fingerprinting~\cite{smith-horwitz}, or the comparison of abstract syntax
trees~\cite{deckard}.
However, all these approaches concentrate on the comparison of code fragments
and not on the comparison of \emph{similar diffs} or commits, as required in our
case.

A diff of a file consists of a sequence of \emph{hunks} that describe the
changes at a textual level.
Every hunk $h$ is introduced by a range information that determines the location
of the changes within a file and contains a section heading $h_\text{head}$.
Section headings display ``the nearest unchanged line that precedes each
hunk''~\cite{diff} and are determined by a regular expression.
Range information is followed by the actual changes: lines $h^+$ that are added
to the new resulting file are preceded by '$+$', lines $h^-$ that are removed
from the original file are preceded by '$-$' and lines $h^\circ$ that did not
change are preceded by a whitespace '\textvisiblespace'.

For the projects considered in the case study, we observed the following
properties:
\begin{itemize}
	\item Commit messages of upstream patches tend to be more verbose, but still
		  are similar to those on patch stacks.
	\item Variable and identifier names do not significantly change between
	      different versions.
	\item Range information of similar hunks changes between different releases.
	\item Section headings tend to stay similar between different releases.
\end{itemize}

In contrast to the detection of code plagiarism or the detection of code
duplicates, in our case the the textual content of diffs between successive
releases of the patch stack tends to stay very close.
For this case, string or edit distances provide an easy but powerful language
independent method for detecting similar code fragments.

Comparing $n$ diffs against each other requires $\mathcal{O}(n^2)$ comparison
operations.
As the necessary string operations are computationally intensive, we employ a
coarse-grained pre-evaluation that serves as a filter:
Two commits can only be similar if both touch at least one common file.
If the intersection of touched files is disjoint the two commits are
automatically considered to be not similar.

Our algorithm calculates a rating for the similarity of the commit message and a
rating for the similarity of the diff.
When comparing diffs, only similar hunks of commonly changes files are compared.
Insertions and deletions are compared independently.

Algorithm~\ref{alg:comp} describes the evaluation of two patches.
The algorithm calculates two ratings, a message rating $r_m\in [0,1]$ and a diff
rating $r_d\in [0,1]$.
$r$ is the weighted arithmetic mean of $r_m$ and $r_d$, weighted by a heuristic
factor $w\in [0,1]$.
If the resulting rating $r < t_i$, the two commit hashes are classified as
dissimilar, if $t_i \leq r < t_a$, then manual evaluation is required, and if $r
\geq t_a$, the commits are classified as similar.
Given a commit hash, \textsc{GetCommit} returns the corresponding message and
diff.
\textsc{StripTags} removes all \emph{tags} (CC:, Signed-off-by:, Acked-by:,
\dots) as they are not relevant for comparing the content of commit messages.
Given the diff of a commit, \textsc{ChangedFiles} returns all touched files of the
diff.
\textsc{GetHunks} returns all hunks of the diff of a file while
\textsc{HunkByHeading} searches for the closest hunk which heading matches $x$
with a rating of at least $t_h$ given a section heading $x$ and the diff of a
file.
\textsc{Dist} takes either two strings or two lists of strings and returns a
rating between 0 and 1, where 0 denotes no commonalities and 1 denotes absolute
similarity.
Our implementation uses the Levenshtein distance, which is a well-known metric
of measuring the similarity of strings.

\begin{algorithm}[h]
	\caption{Detection of similar patches}
	\label{alg:comp}
	\begin{algorithmic}[1]
		\Function{comp}{$a, b, t_a, t_i, t_h, w$}
			\If {not \Call{PreEval}{$a, b$}}
				\State \Return False
			\EndIf
			\State $(\text{msg}_a, \text{diff}_a)\leftarrow$\Call{GetCommit}{$a$}
			\State $(\text{msg}_b, \text{diff}_b)\leftarrow$\Call{GetCommit}{$b$}
			\State $r_m \leftarrow $\Call{Dist}{\Call{StripTags}{$\text{msg}_a$}, \Call{StripTags}{$\text{msg}_b$}}
			\State $r_d \leftarrow []$
			\For {\textbf{each} $\text{file}\leftarrow$\Call{ChangedFiles}{$\text{diff}_a$}}
				\State $\text{hunks}_a\leftarrow$\Call{GetHunks}{diff$_a$, file}
				\State $\text{hunks}_b\leftarrow$\Call{GetHunks}{diff$_b$, file}
				\State $r_f \leftarrow []$
				\For {\textbf{each} $\text{lhunk}\leftarrow\text{hunks}_a$}
					\State $\text{rhunk}\leftarrow$\Call{HunkByHeading}{$\text{hunks}_b, \text{lhunk}_{\text{head}}, t_h$}
					\If {$\text{rhunk}$ is None}
						\State \textbf{continue}
					\EndIf
					\State $r_f$.append(\Call{Dist}{$\text{lhunk}^+, \text{rhunk}^+$})
					\State $r_f$.append(\Call{Dist}{$\text{lhunk}^-, \text{rhunk}^-$})
				\EndFor
				\State $r_d$.append(\Call{Mean}{$r_f$})
			\EndFor
			\State $r_d \leftarrow$\Call{Mean}{$r_d$}
			\State $r\leftarrow w \cdot r_m + (1-w) \cdot r_d$
			\If {$r\geq t_a$}
				\State \Return True
			\ElsIf {$r\geq t_i$}
				\State \Return \Call{InteractiveReview}{$a, b$}
			\EndIf
			\State \Return False
		\EndFunction
	\end{algorithmic}
\end{algorithm}

\section{Discussion}
\begin{figure}
	\centering
	\resizebox{1.0\linewidth}{!}{
		\input{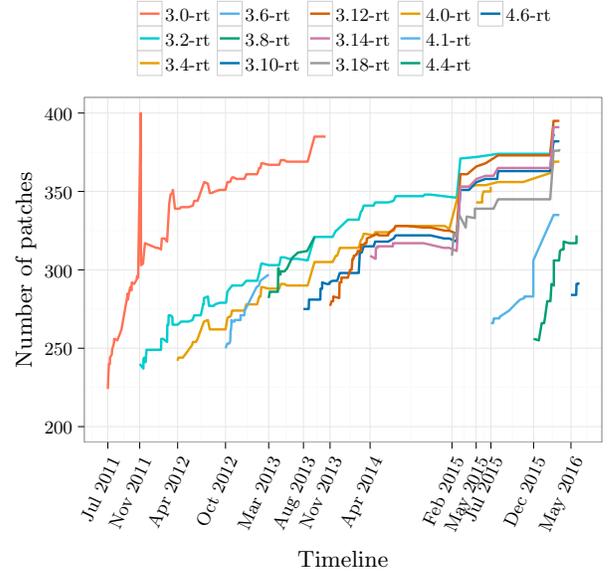}
	}
	\caption{\prt{} patch stack: Evolution of the stack size since Linux kernel
	version 3.0}
	\label{fig:prtevo}
\end{figure}
After grouping all patches into equivalence classes and linking them to optional
commits of the base project, we can distinguish between two temporal conditions:
(1) Patches that first appeared on the patch stack and later appeared in the base
project (ports or \emph{forwardports}) and (2) patches that first appeared in the
base project and were ported back to older versions of the stack (backports).
Patches that are not linked to a commit of the base project are called
\emph{invariant}, as they only appear on the stack.

Across two releases of the patch stack, we observe a flow of patches:
(1) inflow -- new patches on the patch stack and backports.
(2) outflow -- patches that went upstream or patches that were dropped.
(3) invariant -- patches that remain on the stack.

In the follwing, we consider the evolution of the \prt{} patch stack as a case
study:  First, we inspect the temporal evolution of patch stack
size, which is visualised in Figure~\ref{fig:prtevo}.  Among all $554$
releases of the patch stack published since Jule 2011 (that in total consist of
almost $173\,000$ patches), we detected $1042$ different groups of patches.
$195$ of those groups were classified as backports, $153$ groups were classified
as forwardports.

\begin{figure}
	\centering
	\resizebox{1.0\linewidth}{!}{
		\input{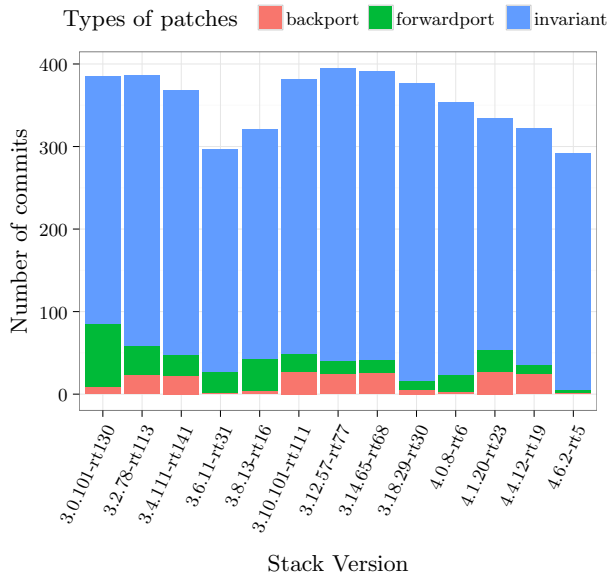}
	}
	\caption{\prt{} patch stack: Comparing the composition of the last major
	releases of the patch stacks}
	\label{fig:prtfuture}
\end{figure}

Knowledge of the stack history allows us to determine the composition
of older patch stacks in terms of the direction of flow of constituents.
Retroactively, we can determine which patches of the stack went upstream at a
later point in time, and compute the amount of backported patches and invariant
patches.
Figure~\ref{fig:prtfuture} shows the composition of the latest releases of major
versions of the \prt{}\cite{prt} patch stack.
Green bars describe the amount of patches on the stack that eventually are
integrated into the upstream code base, red bars describe the amount of
backports, and the blue bars give the number of invariant patches.

\begin{figure}
	\centering
	\resizebox{1.0\linewidth}{!}{
		\input{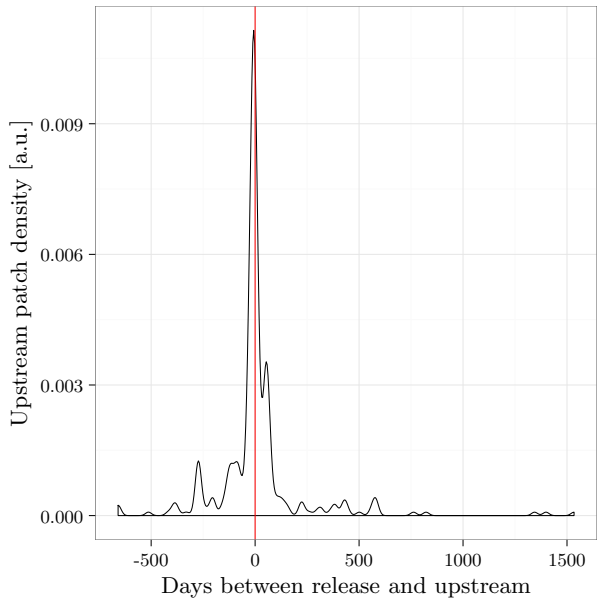}
	}
	\caption{\prt{} patch stack: Distribution of integration times (in days)
	for patches that are eventually integrated in mainline. Positive values
	indicate forwardports, negative values indicate backports.}
	\label{fig:prtups}
\end{figure}

Another covariate of interest is the duration a patch needs to go upstream
(i.e., the time between the first appearance on the patch stack and the
integration with the base project).
Figure~\ref{fig:prtups} shows the result of this analysis for the \prt{}
project.
Positive values on the $x$-axis	describe forwardports, negative values describe
backports.
There is a prominent hot spot around zero days.
We interpret this spot to indicate close cooperation with the base
project: backporting of many patches only takes few days while the
author list of forward and backport patches overlaps.

\section{Conclusions}
We presented an approach and implementation for the quantitative analysis of
patch stacks and a semi-automatic method for identifying similar commits.
An evaluation and visualisation of the \prt{} patch stack was presented as case
study.

In future work, we will concentrate on deeper statistical analysis and
comparing the properties and soft\-ware-en\-gi\-nee\-ring implications of
patch stacks for a various projects.
We are also working on a measure to quantify the invasiveness of patches and
patch stacks, which will allow us to draw conclusions on the eventual
maintenance cost of such stacks.

\bibliographystyle{abbrv}
\bibliography{main}

\end{document}